# Reply to Comment on *Spin-selective reactions of radical pairs act as quantum measurements*

J. A. Jones[a,b], Kiminori Maeda[b,c], U. E. Steiner[d], P. J. Hore[e,*]

[a]Centre for Quantum Computation, Clarendon Laboratory, University of Oxford, Parks Road, Oxford OX1 3PU, UK

[b]Centre for Advanced ESR, University of Oxford, South Parks Road, Oxford OX1 3QR, UK

[c]Department of Chemistry, University of Oxford, Inorganic Chemistry Laboratory, South Parks Road, Oxford OX1 3QR, UK

[d]Fachbereich Chemie, Universität Konstanz, 78465 Konstanz, Germany

[e]Department of Chemistry, University of Oxford, Physical and Theoretical Chemistry Laboratory, South Parks Road, Oxford OX1 3QZ, UK

[*]Corresponding author. Fax: +44 (0)1865 275410.
E-mail address: peter.hore@chem.ox.ac.uk (P.J. Hore).

In his Comment [1] on a recent paper by two of us [2] Kominis claims that the recently proposed quantum measurement model for spin-selective reactions of radical pairs leads to ambiguous predictions in a simple case. Here we show that this claim is based on a confusion between the *unreacted* and *unrecombined* portions of the radical pairs, and to an incorrect interpretation of the improper density matrices used in both our model and the conventional Haberkorn model [3,4] of such reactions. We further show that if this error is corrected then the supposed ambiguity is resolved.

Kominis correctly states that for the simple system he considers, with $\hat{H} = 0$ ($\hat{H}$ is the spin Hamiltonian) and $k_T = 0$ ($k_T$ is the triplet reaction rate), the quantum measurement model predicts that the evolution of the radical pair density matrix is given by

$$\frac{d\hat{\rho}_{RP}}{dt} = -k_S \left( \hat{\rho}_{RP} - \hat{Q}_T \hat{\rho}_{RP} \hat{Q}_T \right). \tag{1}$$

The resulting density matrix is an *improper* density matrix, in that $\text{Tr}(\hat{\rho}_{RP}) \neq 1$ except at $t = 0$. This occurs because, in common with the conventional model, the recombination process is modelled as the simple disappearance of radical pairs, with no corresponding creation of product. This can be overcome by forcibly renormalizing the density matrix, using $\hat{\rho} = \hat{\rho}_{RP} / \text{Tr}(\hat{\rho}_{RP})$, which leads to the evolution equation

$$\frac{d\hat{\rho}}{dt} = -k_S \text{Tr}\left[ \hat{Q}_T \hat{\rho} \hat{Q}_T \right] \left( \hat{\rho} - \frac{\hat{Q}_T \hat{\rho} \hat{Q}_T}{\text{Tr}\left[ \hat{Q}_T \hat{\rho} \hat{Q}_T \right]} \right) \tag{2}$$

in complete agreement with Kominis [1]. Note that the density matrix $\hat{\rho}$ is what is usually considered in the master equations of spin chemistry. Here we note explicitly that it corresponds to the renormalized density matrix of *unrecombined* molecules, which remain in a radical pair state.

Alternatively, the problem can be avoided entirely by expanding the description explicitly to include product terms, and from here on we describe states using this expanded $\{P, S, T\}$ basis. For simplicity we assume that the radical pair begins in the pure spin state

$$\hat{\rho}_0 = |\psi\rangle\langle\psi|, \quad |\psi\rangle = \alpha|S\rangle + \beta|T\rangle \tag{3}$$

with $\alpha\alpha^* + \beta\beta^* = 1$ by normalization, although nothing in our results depends on this assumption. The subsequent evolution according to the quantum measurement model is then given by

$$\hat{\rho}_{\mathrm{QM}}(t) = \begin{pmatrix} \alpha\alpha^*\left[1-e^{-k_St}\right] & 0 & 0 \\ 0 & \alpha\alpha^* e^{-k_St} & \alpha\beta^* e^{-k_St} \\ 0 & \beta\alpha^* e^{-k_St} & \beta\beta^* \end{pmatrix}. \tag{4}$$

The zeros in the top row and left hand column, connecting the radical pair with the reaction products, reflect the fact that coherent superpositions of different chemical species will rapidly decohere, and so can be safely neglected.

This expanded description is now a *proper* density matrix, with $\mathrm{Tr}(\hat{\rho}_{\mathrm{QM}}) = 1$ at all times, and can be rewritten as a sum of two terms:

$$\hat{\rho}_{\mathrm{QM}}(t) = e^{-k_St}\begin{pmatrix} 0 & 0 & 0 \\ 0 & \alpha\alpha^* & \alpha\beta^* \\ 0 & \beta\alpha^* & \beta\beta^* \end{pmatrix} + \left[1-e^{-k_St}\right]\begin{pmatrix} \alpha\alpha^* & 0 & 0 \\ 0 & 0 & 0 \\ 0 & 0 & \beta\beta^* \end{pmatrix}. \tag{5}$$

This description makes clear that the overall density matrix is simply the weighted sum of two constant proper density matrices, with the evolution encoded in the time-varying weights of the *unreacted* and *reacted* terms

$$\hat{\rho}_{\mathrm{QM}}(t) = w_{\mathrm{u}}\hat{\rho}_0 + w_{\mathrm{r}}\left(\alpha\alpha^*|P\rangle\langle P| + \beta\beta^*|T\rangle\langle T|\right). \tag{6}$$

The unreacted component is, as expected, left entirely unchanged by the process, while the reacted component is projected onto a mixture of $|P\rangle\langle P|$ and $|T\rangle\langle T|$ terms, corresponding, respectively, to products formed from radical pairs where the reaction has led to recombination, and to remaining radical pairs where the absence of recombination during the reaction has projected them onto the triplet state [2].

In his Comment Kominis uses a different decomposition of the density matrix, concentrating on the fate of the unrecombined component, which is a combination of the fraction that has not reacted, and the fraction that has been projected onto the triplet state by the reaction. In general one must be cautious in ascribing significance to particular decompositions of a density matrix in order to avoid falling into the *Partition Ensemble Fallacy* [5], but such a decomposition can be made, giving

$$\hat{\rho}_{\mathrm{QM}}(t) = \begin{pmatrix} 0 & 0 & 0 \\ 0 & \alpha\alpha^* \mathrm{e}^{-k_{\mathrm{S}}t} & \alpha\beta^* \mathrm{e}^{-k_{\mathrm{S}}t} \\ 0 & \beta\alpha^* \mathrm{e}^{-k_{\mathrm{S}}t} & \beta\beta^* \end{pmatrix} + \left[1 - \mathrm{e}^{-k_{\mathrm{S}}t}\right] \begin{pmatrix} \alpha\alpha^* & 0 & 0 \\ 0 & 0 & 0 \\ 0 & 0 & 0 \end{pmatrix}. \tag{7}$$

The use of an unnatural decomposition leads to component matrices which are not proper density matrices (they do not have trace equal to one) and one of which varies with time. This time-dependence reflects the fact that the unrecombined component includes varying proportions of the unreacted component and *part* of the reacted component, turning gradually from a superposition of singlet and triplet states into a triplet-only state as the singlet component is converted to product. No particular significance can be ascribed to the varying trace, as this matrix only describes an arbitrarily selected component of the density matrix.

Kominis then uses this decomposition to find the normalized density matrix describing the unrecombined radical pairs, arguing that

$$\hat{\rho}(t) = w_0 \hat{\rho}_0 + (1 - w_0)|\mathrm{T}\rangle\langle\mathrm{T}| \tag{8}$$

with $w_0 = \mathrm{e}^{-k_{\mathrm{S}}t}$ and the size of the second term determined by imposing normalization, and he claims that this approach leads to an inconsistent equation for $\hat{\rho}(t)$. His calculations, however, use incorrect forms for the two weights, and when these are corrected the inconsistency is resolved. The full decomposition, equations (5) and (6), shows that

$$\hat{\rho}_{\mathrm{QM}}(t) = w_0 \hat{\rho}_0 + w_{\mathrm{T}}|\mathrm{T}\rangle\langle\mathrm{T}| + w_{\mathrm{P}}|\mathrm{P}\rangle\langle\mathrm{P}|. \tag{9}$$

and it is the sum of these three weights which must be normalised, so that $w_0 + w_{\mathrm{T}} + w_{\mathrm{P}} = 1$; the forms used by Kominis normalize $w_0 + w_{\mathrm{T}}$, incorrectly assuming that $w_{\mathrm{P}} = 0$ throughout the reaction. Correcting this error gives

$$\hat{\rho}(t) = \frac{w_0 \hat{\rho}_0 + w_{\mathrm{T}}|\mathrm{T}\rangle\langle\mathrm{T}|}{w_0 + w_{\mathrm{T}}} \tag{10}$$

which leads [6] to precisely the same result as equation (2).

We note in passing that the dangers inherent in the use of improper density matrices are not, of course, confined to the quantum measurement model, and very similar behaviour is seen with the conventional Haberkorn approach, for which the density matrix is

$$\hat{\rho}_{\mathrm{HK}}(t) = \begin{pmatrix} 0 & 0 & 0 \\ 0 & \alpha\alpha^* \mathrm{e}^{-k_{\mathrm{S}}t} & \alpha\beta^* \mathrm{e}^{-k_{\mathrm{S}}t/2} \\ 0 & \beta\alpha^* \mathrm{e}^{-k_{\mathrm{S}}t/2} & \beta\beta^* \end{pmatrix} + \left[1 - \mathrm{e}^{-k_{\mathrm{S}}t}\right] \begin{pmatrix} \alpha\alpha^* & 0 & 0 \\ 0 & 0 & 0 \\ 0 & 0 & 0 \end{pmatrix}. \tag{11}$$

In the Haberkorn model, however, no distinction is possible between the unreacted and unrecombined components, and in this case the use of time-varying density matrices is inevitable.

In conclusion, we have resolved the supposed ambiguity in the quantum measurement model of spin-selective reactions, showing that the theory is indeed self-consistent. This does not, however, resolve the question of whether the model is correct, and arguments of this kind [7] must be based on other factors.